 \documentclass{aa}
\usepackage{graphicx}
\usepackage[sectionbib,authoryear]{natbib}

\usepackage{times}
\usepackage{graphicx} 
\newcommand{\gr}{$\gamma$-ray \,}
\newcommand{\grs}{$\gamma$-rays \,}
\begin{document}

\title{Internal dynamics and particle acceleration in Tycho's SNR}
 
\titlerunning{Internal dynamics and particle acceleration in Tycho's SNR}
\authorrunning {V\"olk et al.} 

\author{H.J. V\"olk \inst{1}
 \and E.G. Berezhko \inst{2} 
 \and L.T. Ksenofontov \inst{2}
}

\institute{
  Max-Planck-Institut f\"ur Kernphysik, P.O. Box 103980, D 69029
  Heidelberg, Germany\\
              \email{Heinrich.Voelk@mpi-hd.mpg.de}
  \and
  Yu.G. Shafer Institute of Cosmophysical Research and Aeronomy,
                     31 Lenin Ave., 677980 Yakutsk, Russia\\
              \email{berezhko@ikfia.ysn.ru; ksenofon@ikfia.ysn.ru}
}

\offprints{H.J.V\"olk}

\abstract {} 
{The consequences of a newly suggested value for the SN explosion
energy $E_\mathrm{sn} = 1.2 \times 10^{51}$~erg are explored for the case of
Tycho's supernova remnant (SNR).}  {A nonlinear kinetic theory of cosmic ray
(CR) acceleration in supernova remnants (SNRs) is employed to investigate the
properties of Tycho's SNR and their correspondence to the existing experimental
data.}  
{It is demonstrated that the large mean ratio between the radii of the contact
discontinuity and the forward shock is consistent with the very effective
acceleration of nuclear energetic particles at the forward shock.  It is also
argued that consistency of the value $E_\mathrm{sn} = 1.2 \times 10^{51}$~erg
with the gas dynamics, acceleration theory, and the existing \gr measurements
requires the source distance $d$ to be greater than $3.3$~kpc. The
corresponding ambient gas number density is lower than $0.4$~cm$^{-3}$. Since
the expected \gr flux strongly depends on the source distance,
$F_{\gamma}\propto d^{-7}$, a future experimental determination of the actual
\gr flux from Tycho's SNR will make it possible to determine the values of the
source distance $d$ and of the mean ambient gas density. A simple inverse
Compton model without a dominant population of nuclear CRs is not compatible
with the present upper limit for the \gr emission for any reasonable ambient
interstellar B-field.}  
{Given the consistency between acceleration theory and overall, as well as
internal, gas dynamics, a future \gr detection would make the case for {\it
nuclear} \rm particle acceleration in Tycho's SNR incontrovertible in our
view.}


\maketitle

\section{Introduction}
Nonlinear kinetic theory of diffusive CR acceleration in SNRs \citep{byk96,
bv97} couples the gas dynamics of the explosion with the particle
acceleration. The present form of the solution assumes spherical
symmetry. In this approximation it is possible to predict the
evolution of gas density, pressure, and mass velocity, as well as the radii of
the forward shock and the contact discontinuity, together with the energy
spectrum and the spatial distribution of CR nuclei and electrons, including the
properties of their non-thermal radiation. Applied to individual young SNRs
\citep[see][for reviews]{vlk04, ber05, ber08} this theory has successfully
explained many observed SNR properties. Making use of the observed synchrotron
emission spectrum from radio to X-ray frequencies it permits the derivation of
the injection rate of nuclear particles, essentially that of protons, into the
acceleration process. It has also allowed the determination of the degree of
magnetic field amplification, a process advocated earlier from plasma
simulations \citep{lb00,belll01}. The initial examples of such a systematic
analysis concerned SN 1006 \citep{bkv02} and Tycho's SNR
\citep{vbkr}\footnote{First indications for high magnetic field strengths $B
= 10^2 - 10^3 \mu$G had been obtained much earlier for Tycho's SNR by
\citet{re92} from an analysis of the radio spectrum alone.}.
It was specifically predicted that magnetic field amplification leads to
the concentration of the highest-energy electrons and their corresponding
synchrotron and IC \gr emission in a very thin shell just behind the forward
shock \citep[see Fig.7 of][]{bkv02}. This concentration, which is a result of
synchrotron cooling, is observationally broadened due to a projection effect
\citep{bkv03}\footnote{Such filamentary structures have been observed
in hard X-rays \citep{vl03,long03,bamba03} and are nowadays used as a second
independent method to infer the magnitude of the amplified field. As it should
be, both methods lead to consistent results within the errors, whenever both
data sets exist \citep{vl03,bv04b,vbk05,ballet06,par06}.}.

As indicated above, the theory has been used in some detail to investigate
Tycho's SNR (G120.1+1.4) as the remnant of a type Ia SN in a (roughly) uniform
interstellar medium (ISM), in order to compare the results with existing
data. A stellar ejecta mass $M_\mathrm{ej}=1.4M_{\odot}$, distance $d=2.3$~kpc,
and ISM number density $N_\mathrm{H}=0.5$~ H-atoms cm$^{-3}$ \citep{vbkr,vbk05}
were used. For these parameters a total hydrodynamical explosion energy
$E_\mathrm{sn}=0.27\times 10^{51}$~erg was originally derived to fit the
observed size $R_\mathrm{s}$ and expansion speed $V_\mathrm{s}$.

The steep and concave radio synchrotron spectrum shows that the forward shock
is characteristically modified by the pressure of the accelerated nuclear
particle component, with the weaker subshock essentially accelerating the radio
electrons, while higher-energy electrons increasingly ``see'' the overall shock
transition \citep{re92, bkv02}.

Since the hardening of the particle momentum spectrum begins at momenta $p
\approx m_\mathrm{p} c$, the frequency range, where the corresponding hardening
of the spatially integrated synchrotron spectrum occurs, brackets the effective
mean magnetic field strength $B_\mathrm{d}$ inside the SNR; here $m_\mathrm{p}$
denotes the proton mass. The magnetic field strength is in addition constrained
by the requirement that the calculated electron spectrum leads to a smooth
synchrotron cutoff whose frequency agrees with the X-ray observations. From
such a comparison of the calculated and observed overall synchrotron spectra
the field strength resulted as $B_\mathrm{d} \approx 240 \mu$G. Assuming the
mean magnetic field strength in the circumstellar medium of Tycho's SN to be
equal to a typical ISM-value of $5 \mu$G, and using an overall shock
compression ratio $\sigma < 6$~(see below), pure MHD-compression would give a
much smaller interior field strength $B_\mathrm{d} < 30 \mu$G\footnote{Since
injection of nuclear particles occurs only at quasi-parallel shocks
\citep{vbk03}, a pure MHD-compression of a $5 \mu$G upstream field would lead
to even considerably lower interior field strengths $\sim 10 \mu$G in an
unmodified shock with a compression ratio of 4. This property will be used in
subsection 3.1.}.  A later re-analysis, including now independent Chandra
observations of thin X-ray filaments as indicators of the outer shock, gave
$B_\mathrm{d} \approx 300 \pm 60 \mu$G \citep{vbk05, vbk05a}. Overall, together
with the approximation for the scattering properties of the magnetic field
fluctuations -- summarized in the next section -- good consistency of the
predictions of the nonlinear theory with the existing observational data was
achieved.

Recently the radius $R_\mathrm{c}$ of the contact discontinuity (CD),
separating the shocked ISM and the ejecta material, has been investigated by
\citet{warren05}, using {\it Chandra} X-ray observations. The ratio
$R_\mathrm{c}/ R_\mathrm{s}$ of $R_\mathrm{c}$ and the radius $R_\mathrm{s}$ of
the forward shock (blast wave) is a new physical
variable characterizing Tycho's SNR. It connects the {\it internal} dynamics
with the particle acceleration process. The large mean value
$R_\mathrm{c}/ R_\mathrm{s}=0.93$ found for this ratio was
interpreted as evidence for efficient CR acceleration which makes the medium
between those two discontinuities considerably more compressible. Without
accelerated particles, and in a spherically symmetric calculation, the value of
the ratio is significantly smaller, $R_\mathrm{c}/ R_\mathrm{s}=
0.77$~\citep{wang01}. We have demonstrated \citep{vbk07, vbk08} that the
observed ratio $R_\mathrm{c}/ R_\mathrm{s}=0.93$ agrees quite well with the
results of \citet{vbkr}.

Somewhat later \citet{bad06} published a detailed comparison between the
high-quality X-ray observations from XMM-Newton and Chandra of the ejecta
emission and current models for Type Ia explosions. They found that the
fundamental properties of the X-ray emission in Tycho are well reproduced by a
one-dimensional delayed detonation model with a kinetic energy $E_\mathrm{sn} =
1.2\times 10^{51}$~erg. Such a value is compatible with well-known results for
deflagration models \citep{nomoto} as well. In any case, this explosion energy
is rather in the middle of the typical range of type Ia SN explosion energies
that vary by a factor of about two \citep[e.g.][]{gam04,brs06}. In our initial
paper \citep{vbk07} we had indicated that also such a higher kinetic energy of
the explosion is consistent with the SNR dynamics.

Here we shall a priori adopt the value $E_\mathrm{sn}= 1.2\times 10^{51}$~erg
and perform the calculation of all relevant physical properties of Tycho's SNR
connected with nonthermal energy production and emission in order (i) to find
out how well the existing SNR data -- especially the ratio $R_\mathrm{c}/
R_\mathrm{s}$ -- are consistent with such an explosion energy (ii) to
investigate the expected ratio of leptonic to hadronic \gr emission, and (iii)
to possibly determine the allowed range for the source distance. In section 2
we shall outline our approximation for the scattering properties of the
magnetic field fluctuations, whereas section 3 contains the results and the
discussion.

\section{Approximation for the scattering properties of the system}

Recently \citet{bell04} found that a non-resonant streaming
instability will occur within the precursor of a strong, accelerating
shock. The diffusive streaming of accelerated CRs is expected to be so
strong in this region that a purely growing MHD mode appears with a
growth rate that is, at least at the beginning of the precursor,
larger than the growth rate of the well-known resonant Alfv\'enic
mode. It is expected that due to this non-resonant instability the
external magnetic field $B_\mathrm{ISM}$ is amplified within the
entire precursor structure, since the main effect is produced by the
most energetic CRs which populate the whole precursor diffusively
during their acceleration. The
Alfv\'en wave excitation in the shock precursor
\citep{bell78,bo78,mv82} corresponds to an additional unstable mode
that in particular leads to a high level of resonantly
scattering waves. This mode has been argued to dominate the {\it
magnetic field energy deep in the precursor}, after saturation of the
non-resonant instability \citep{pell06}. Therefore the overall
field amplification will be the result of both instabilities operating
in the precursor.

The saturation value $B_\mathrm{nr}$ of the non-resonantly amplified
magnetic field is given by \citep{bell04,pell06}

\begin{equation}
\frac{B_\mathrm{nr}^2 }{8\pi}\sim \frac{3~ V_\mathrm{s}~P_c}{2~ \Phi~ c} 
\approx 1.5 \times 10^{-3}~P_\mathrm{c}~
\left({V_\mathrm{s}\over {3000~
\mathrm{km/s}}}\right),
\label{eq1}
\end{equation}
where $P_\mathrm{c}$ is the resulting CR pressure,
$\Phi=\log(\epsilon_\mathrm{max}/m_\mathrm{p}c^2)$, $\epsilon_\mathrm{max}$ is
the maximal CR energy, $m_\mathrm{p}$ is the proton mass, 
and the value $\Phi=10$ was
adopted in the above relation.
This value of $B_\mathrm{nr}^2 /(8\pi)$ is by a factor $ \approx 3.3$
smaller than the value 
\begin{equation}
B_0^2 /(8\pi) \approx 5\times 10^{-3} P_\mathrm{c},
\label{eq2}
\end{equation}
empirically determined for several SNRs, where the upstream (amplified)
magnetic field strength $B_0$ in the shock precursor, defined as $B_0 =
B_\mathrm{d}/\sigma$, has been derived both from the spatially integrated radio
and X-ray synchrotron spectrum, as well as from the thickness of the X-ray
filaments which determine the downstream field strength $B_\mathrm{d}$
\citep{bv06}.

The subsequent growth of the resonant Alfv\'en instability towards the
precursor takes place in a medium where the effective Alfv\'en
velocity is given by $c_\mathrm{A}= B_\mathrm{nr}/\sqrt{4\pi
\rho}$ and $\rho$ is the mass density. 
Following \citet{pell06} (see their section 3.2.2.) 
one obtains at the subshock
\begin{equation}
\frac{B_\mathrm{res}^2 }{ B_0^2} \approx 7~ 
\left({V_\mathrm{s}\over {3000~
\mathrm{km/s}}}\right)^{1/2}.
\label{eq3}
\end{equation}
This expression shows that the existing mechanisms are able to amplify the
magnetic field within the precursor to a strength, 
which considerably exceeds the
required value (2) \citep{ber08}. 
In fact, it would be more appropriate to use in Eq.(3) instead of $B_0$
the value
$B_1=(\sigma/\sigma_\mathrm{s})B_0$, which in our model is the field strength
just ahead of the subshock,
where $\sigma_\mathrm{s}$ is the subshock compression ratio.
This reduces the numerical factor 7 on the r.h.s. of Eq.(3) to about 4
for an object like Tycho's SNR,
which still substantially exceeds the required value.
However,
this contribution $B_\mathrm{res}^2 /(8\pi)$ of the
resonant instability to the magnetic field energy density disregards
any nonlinear wave dissipation. Indeed, within the framework of weak
plasma turbulence theory, the resonant instability is growing freely
up to the subshock, where the growth rate goes to zero abruptly.

However, the MHD simulations by \citet{bell04}, recently confirmed with higher
numerical resolution by \citet{Zirakashvili08}, show strong dissipation
of the excited {\it non-resonant} instability into the thermal plasma
by induced gas compressions which dissipate in shocks. At the same time
the field strength is expected to reach a value $B_\mathrm{nr}
\gg B_\mathrm{ISM}$, discussed above, at which the instability
saturates \citep[for a different view, see][]{np07}. It is true that
these simulations do not directly describe the situation in the
nonuniform shock precursor with its mixture of non-resonant and
resonant fluctuations, where closer to the subshock more and more
resonant particles appear. Nevertheless the shock dissipation of
transverse magnetic turbulence into the thermal gas is likely to be a
generic feature for the mixture of non-resonant and resonant
instability modes, and should in particular also effectively dissipate
the developing resonant waves.  

This theoretical expectation of strong nonlinear wave dissipation is
consistent with Eq.(2) which shows that the uninhibited growth of the
resonant instability leads to a substantial overestimate for the
magnetic wave energy density at the subshock. It is therefore a
reasonable approximation that at some level of magnetic field
amplification beyond non-resonant saturation the nonlinear dissipation
by shock formation balances the linear growth of the resonant
instability, the remaining energy going into thermal energy of the
precursor plasma.

The spectrum of CRs produced by a strong, modified shock is very hard
so that the CRs with the highest energies make the largest
contribution to the overall CR energy density. These most energetic
CRs produce a field amplification on their own spatial scale which is
the precursor size.  Therefore CRs with lower energies already ``see''
the amplified field $B_0$ as a mean field. Beyond that the strongly
excited resonant waves lead to strong particle scattering that
approaches the Bohm limit, where the scattering mean free path locally
equals the gyro radius in the amplified field $B$.

Since the process of magnetic field amplification is not included in
our kinetic theory we postulate the existence of a far upstream
amplified field $B_0 \gg B_\mathrm{ISM}$ that is determined from
observations of the downstream magnetic field strength $B_\mathrm{d}$,
and a simplified connection
\begin{equation}
B = (\rho / \rho_\mathrm{d})~B_\mathrm{d}
\label{eq4}
\end{equation}
between the local field strength $B$ and the local density $\rho$.
Together with the above equation for $B_0$, this defines the amplified
magnetic field everywhere in the shock transition and is used in our
models for the particle acceleration and
gamma-ray emission \citep[e.g.][]{ber08}.

A value $B_0$ of the order of $10~ B_\mathrm{ISM}$ or even larger, as
derived in the present paper \citep[and for other sources:][]{vbk05,
ber05, ber08}, appears quite plausible for young objects, in which
longer-term dissipation mechanisms \citep{pz03,pz05} have not yet had
sufficient time to operate.

The problem is ultimately the scattering (or better, the scattering
strength) of the highest-energy particles at the instantaneous
cutoff. This is an intrinsically time-dependent process that is not
yet resolved in particle acceleration theory even for unmodified
shocks \citep[e.g.][]{lc83}, let alone for nonlinearly modified shocks
\citep{ab06}.

This interpretation is different from that of \citet{pz03,pz05} who
were not yet aware of the non-resonant instability. It is also
different from that of \citet{ab06} and \citet{veb06} who do not
include the non-resonant instability and its field amplification
effect in their arguments, and rather concentrate on the nonlinear
effects of the resonant Alfv\'en wave instability in the spirit of
\citet{belll01}. It is finally also different from the recent
calculation of the acceleration properties of shocks by
\citet{ZirakPtuskin08} who base their work entirely on the
non-resonant instability and its amplification properties.

Our approximation for young objects consists then first of all in
assuming that the magnetic field strength is amplified everywhere
around the shock in the sense that {\it all} accelerating particles
experience the amplified field in the region upstream of the
subshock. The actual strength of $B_0$ is phenomenologically
determined by iteratively fitting the theoretically calculated
electron spectrum to the observed synchrotron spectrum, and/or by the
analysis of observed X-ray filaments \citep [e.g.][]{vbk05}. The
effective downstream field strength $B_{\mathrm{d}} = \sigma B_0$ is
the result of the MHD-compression of this largely perpendicular
upstream field. On account of the large amplitude $\langle~\delta
B^2~\rangle ~\sim B^2$ of the resonantly scattering waves, the
diffusion coefficient is taken as the Bohm limit in the local
amplified field $B = (\rho / \rho_d)B_\mathrm{d}$. This is clearly a
lower limit to the scattering mean free path \citep[e.g.][]{marc06},
especially for the highest-energy particles whose maximum energy is
almost certainly overestimated in this way. However the Bohm limit
seems to be roughly consistent with the observations \citep{par06}.

Secondly, we assume the dissipation rate of magnetic field energy into
the thermal gas to be equal to the linear growth rate~ $c_\mathrm{A}
\nabla P_\mathrm{c}$ of the resonant Alfv\'enic wave field, with~
$c_\mathrm{A}= B/\sqrt{4 \pi \rho}$ \citep [see also][]{ber08}. This
does not directly include the dissipative limitation of the growth of
the non-resonantly unstable modes demonstrated by the MHD simulations
and to that extent still underestimates the overall gas heating. The
contribution of the non-resonant modes to the gas heating is only
taken into account indirectly through the enhancement of the Alfv\'en
velocity in the dissipation term.

For Tycho's SNR, as for all other cases studied, this gas heating lowers the
overall compression ratio to values of the order of 5 to 7 which seem very
plausible numbers in the light of the properties of the synchrotron emission
.

It is clear from these arguments that the complexity of the magnetic
field, even in the asymptotic case of strong shocks, can at present
only be partially resolved from plasma theory alone. The {\it
observed} synchrotron emission properties of SNRs however determine
our key parameter $B_0$ which we use
to give an approximate description as outlined above. This does not
yet address the questions of the time dependence of this quantity (due
to the higher shock velocity in the past) or of the late-time escape of
high-energy particles. In this paper we shall assume that $B_0$ is a
constant and fixed by the measurements at the present time. We note however
that even a simplified picture with time-independent upstream 
field $B_0$ contains such an important element
as escaping CRs \citep{bek88,byk96} which are the highest-energy CRs
whose outward diffusion is faster than the SNR shock expansion.
In reality one has to expect this loss process to be
even much faster due to two factors. First of all, a decreasing
$B_0(t)$ due to the shock deceleration makes high-energy CR diffusion
progressively faster. Secondly, these escaping CR particles that exist
far upstream from the SN shock produce less efficiently magnetic field
perturbations and therefore their diffusion coefficient becomes
progressively closer to the average diffusion coefficient in the ISM.

\section{Results and Discussion}

The distance determinations for Tycho's SNR have varied in the past.
\citet{dev85} gave a most likely range as $d = 3.2 \pm 0.3$~kpc. \citet{alb86}
found $2.2^{+1.5}_{-0.6}$~kpc, whereas \citet{sch95} argued that $d$ could be
as large as 4.5 kpc. Using $\mathrm{H}_{\alpha}$ measurements and assuming a
strong gas shock with a compression ratio of $\sigma=4$ \citet{kwc87} derived
distances of 2.0--2.8 kpc, and \citet{s91} extended this to 1.5--3.1 kpc. Very
recently \citet{r-l04} estimated $d=2.83 \pm 0.79$~kpc by comparing Tycho's SNR
to SN~1006, and $d=2.85 \pm 0.4$~kpc by using the peak luminosity-decline
correlation for type Ia Supernova explosions. With the exception of the upper
limit of \citet{sch95}, all these estimates lie below about 3.6 kpc. In such a
situation we explore here the range of the distances $d=3.1-4.5$~kpc. As the
lowest value $d=3.1$~kpc is taken because distances $d\leq 3.1$~kpc are
inconsistent with existing TeV \gr measurements, as shown below.

For any given value $d$ we find the density $N_\mathrm{H}$ of the ambient ISM
from a fit to the observed SNR size and expansion speed. Due to the relatively
small uncertainties of the measurements of angular radius
$R_\mathrm{s}/d$ and expansion speed $V_\mathrm{s}/d$, there exists an almost
unique value of $N_\mathrm{H}$ for every given distance $d$, given the
explosion energy $E_\mathrm{sn}$.

\begin{figure}
\centering
\includegraphics[width=7.5cm]{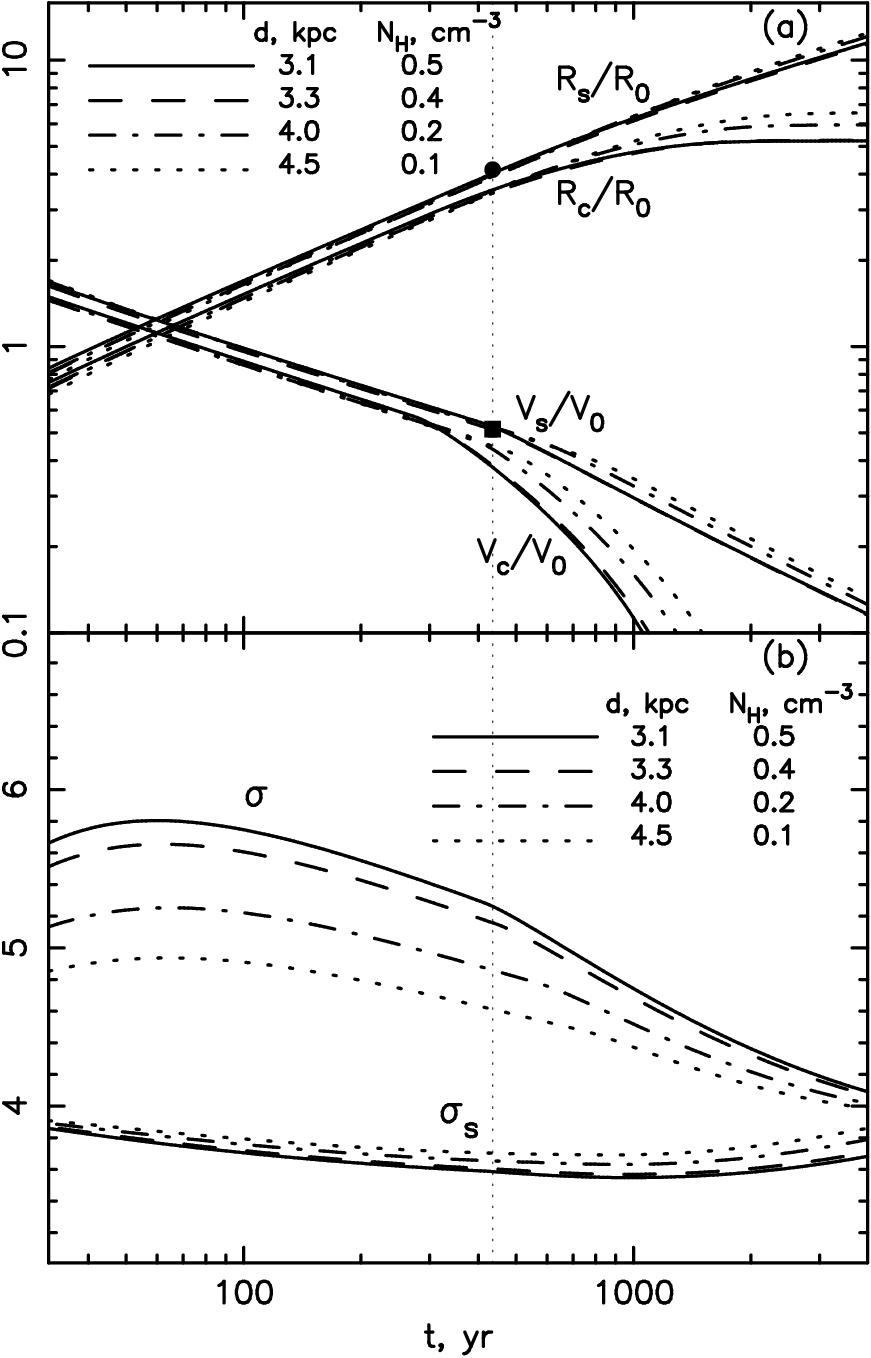}
\caption{(a) Shock (contact discontinuity) radius $R_\mathrm{s}$
  ($R_\mathrm{c}$) and shock (contact discontinuity) speed $V_\mathrm{s}$
  ($V_\mathrm{c}$) in units of $R_0=(4 \mbox{kpc}/d)$~pc and $V_0=10^4(4
  \mbox{kpc}/d)$~km/s, (b) total shock ($\sigma$) and subshock
  ($\sigma_\mathrm{s}$) compression ratios for Tycho`s SNR as functions of
  time, calculated for four different distances $d$. The {\it dotted vertical
  line} marks the current epoch, $t=435$~yrs. The observed mean size ({\it
  filled circle}) and speed ({\it filled square}) of the shock, as determined
  by radio measurements \citep{tg85}, are shown as well.}
\label{f1}
\end{figure}

Fig.\ref{f1} shows the calculations of shock-related hydrodynamic quantities
for four different source distances in the range $3.1\leq d\leq 4.5$~kpc,
together with the azimuthally averaged experimental data from radio
observations \citep{tg85}. The corresponding ambient gas densities, which
result from a fit to the observed angular SNR size and expansion speed, are
listed in Table~1 together with the values of other relevant parameters at the
current epoch.

The system is just in the transition from free expansion into the Sedov
phase. Therefore, although inevitably rough, an analytical explanation of the
resulting relation between $N_\mathrm{H}$ and $d$ is the following: since  
in the Sedov phase $R_\mathrm{s}\propto
(E_\mathrm{sn}/N_\mathrm{H})^{1/5}$, the density $N_\mathrm{H}\propto
E_\mathrm{sn}/d^5$ strongly decreases with increasing distance $d$.

As in our previous studies \citep{vbkr,vbk05,vbk07} we determine the values of
the proton injection rate $\eta$, electron-to-proton ratio $K_\mathrm{ep}$ and
interior magnetic field value $B_\mathrm{d}$ from a fit to the observed
synchrotron spectrum.  The softness of the observed low-energy radio spectrum
-- relative to a test particle spectrum -- requires a proton injection rate
$\eta=3\times 10^{-4}$ in all cases for different distances.  This implies a
significant nonlinear modification of the shock at the current age of
$t=435$~yrs (see Fig.\ref{f1}b).

\begin{figure}
\centering
\includegraphics[width=7.5cm]{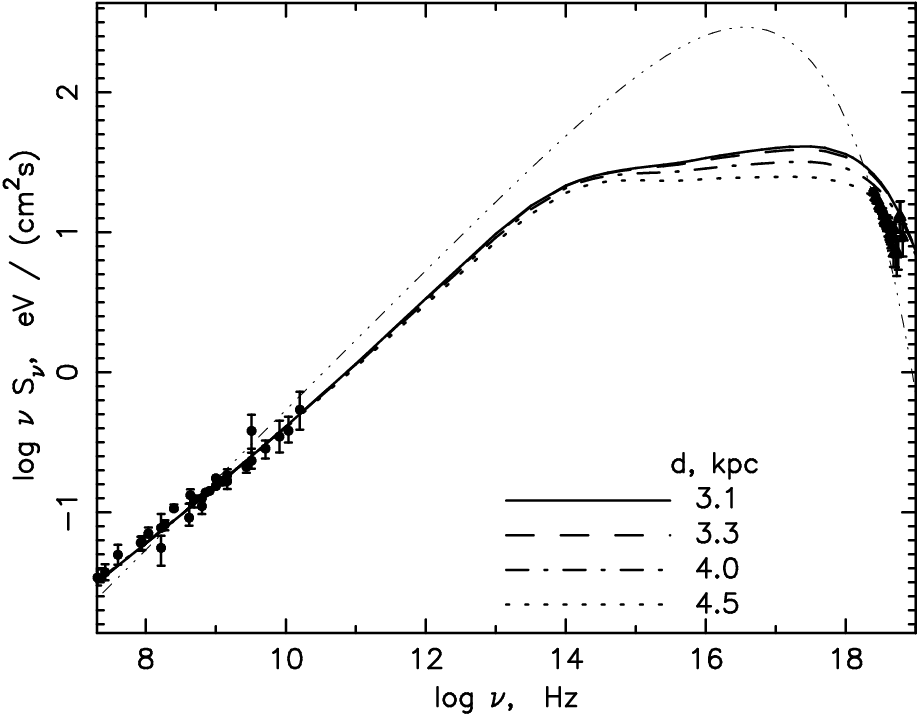}
\caption{Synchrotron emission flux as a function of frequency for the
same four cases as in Fig.\ref{f1}, at the current time. 
The {\it thin dot-dot-dashed line} represents the flux expected in the test
particle limit for an interior magnetic field strength
$B_\mathrm{d}=20$~$\mu$G.
The observed X-ray \citep{agp99} and radio
emission
\citep{re92} are shown as well.} 
\label{f2}
\end{figure}

To fit the measured synchrotron flux a large magnetic field
$B_\mathrm{d}\sim 400 \mu$G is required (see Table~1). Only such a high
magnetic field strength can provide sufficiently large synchrotron losses to
fit the observed X-ray flux. As Fig.\ref{f2} shows, one then can obtain a fit
to the synchrotron radio and X-ray spectra which is as good as in our previous
studies \citep{vbkr, vbk05}, which assumed $E_\mathrm{sn}=0.27\times
10^{51}$~erg and in \citet{vbkr} also a slight overestimate of the observed
X-ray emission.

\begin{table*}
\caption{Models Parameters (Tycho SNR)}             
\label{table1}      
\centering                          
\begin{tabular}{ c | c c c c c c c}        
\hline\hline                 

 $d$, kpc
      & $N_\mathrm{H}$, cm$^{-3}$
             & $\sigma$
                   & $B_\mathrm{d}$, $\mu$G
                         & $B_\mathrm{d}'$, $\mu$G
                               & $K_\mathrm{ep}$, $10^{-4}$
                                     &$F_{\gamma}^{\pi}/F_{\gamma}^{IC}$ \\  
\hline                        
 3.1  & 0.5  & 5.25 & 420 & 360 & 4.2 & 40  \\
 3.3  & 0.4  & 5.15 & 412 & 350 & 5.1 & 27 \\
 4.0  & 0.2  & 4.85 & 388 & 320 & 9.5 & 7.7  \\
 4.5  & 0.1  & 4.6  & 368 & 309 & 15. & 2.4  \\
\hline                                   
\end{tabular}
\end{table*}

\begin{figure}
\centering
\includegraphics[width=7.5cm]{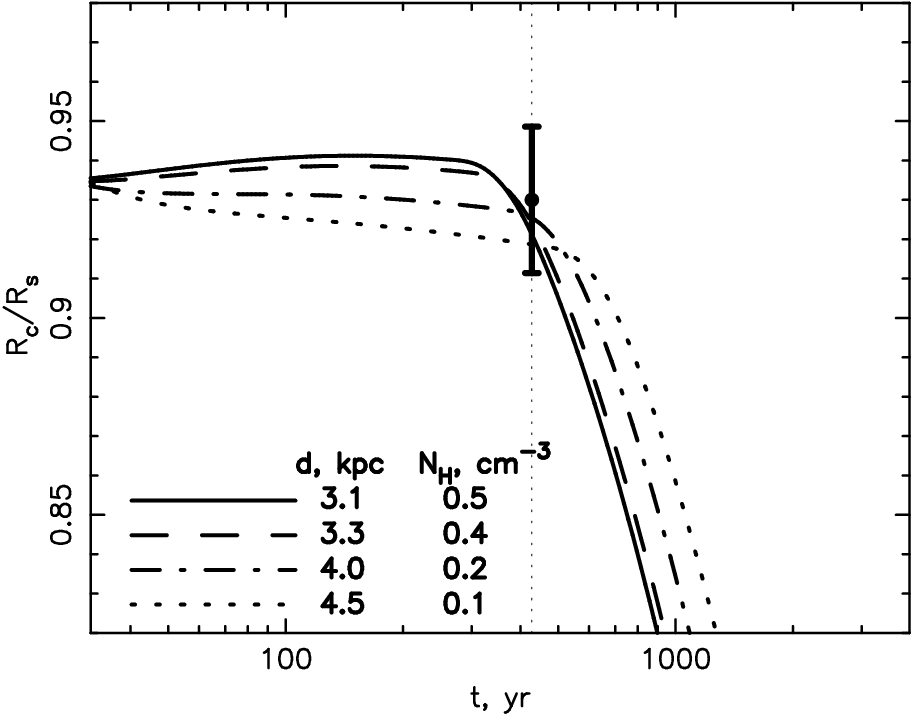}
\caption{The ratio $R_\mathrm{c}/R_\mathrm{s}$ of the contact
discontinuity radius to the forward shock radius
as a function of time, calculated for four different source distances, together
with the
observationally determined value of \citet{warren05}.}
\label{f3}
\end{figure}

We now compare the theoretical ratio $R_\mathrm{c}/R_\mathrm{s}$ of the
contact discontinuity radius (CD) and the blast wave radius with the
observational estimate $R_\mathrm{c}/R_\mathrm{s} = 0.93$ by
\citet{warren05}. In order to do this, one has to first make the two results
comparable. It was argued in \citet{vbk07,vbk08} that if one starts from a
spherically symmetric calculation of the CD radius as we do, one has to take
into account that the actual CD is subject to the Rayleigh-Taylor 
instability, and thus a correction is needed to compare such a 1-D calculation
with observations of the CD. Here we also correct our calculated CD radius by a
factor 1.05. It is clear that the increased
compressibility of the shocked material can only be due to accelerated nuclear
particles, because the pressure of the accelerated electrons is always
negligibly small.

The comparison of the corrected values $R_\mathrm{c}/R_\mathrm{s}$ with the
experimentally estimated value $R_\mathrm{c}/R_\mathrm{s}=0.93$ (in
Fig.\ref{f3} we present that value with 2\% uncertainties, according to
\citet{warren05}) shows quite good agreement in all the cases considered (see
Fig.\ref{f3}). According to the theoretical calculation the ratio
$R_\mathrm{c}/R_\mathrm{s}$ is almost constant in the free expansion phase of
the SNR evolution, whereas in the adiabatic phase the contact discontinuity
quickly drops behind the forward shock and the ratio starts to
decrease. According to Fig.\ref{f1} and Fig.\ref{f3} Tycho's SNR is in transit
to the adiabatic phase. In the case without CR acceleration our corrected value
is $R_\mathrm{c}/R_\mathrm{s} \approx 0.89$. We note that this value is larger
(and the effect of CRs correspondingly smaller) than that assumed by
\citet{warren05}, because these authors referred to SNR parameters according to
which Tycho's SNR is closer to the adiabatic phase \citep{wang01}.

\begin{figure}
\centering
\includegraphics[width=7.5cm]{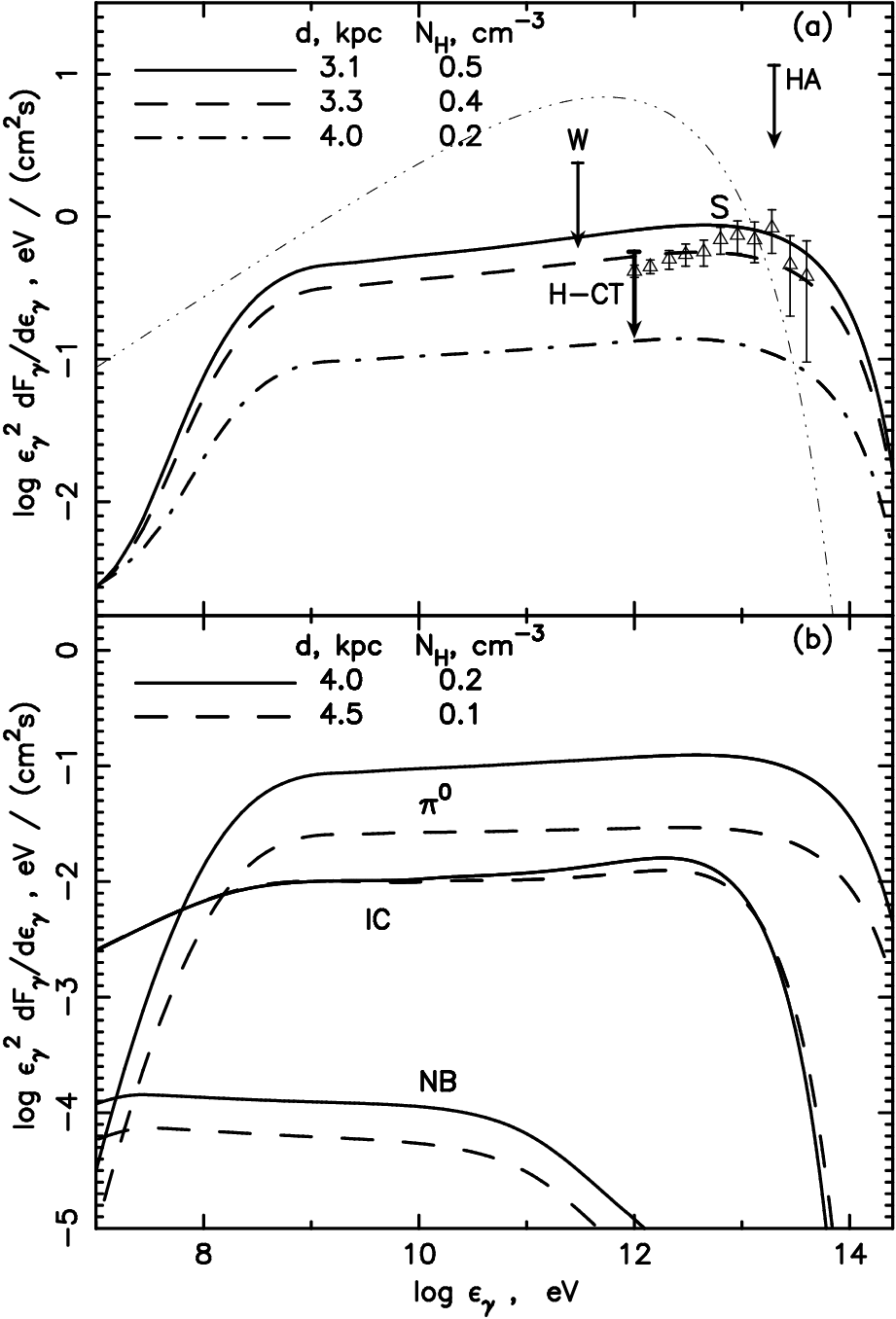}
\caption{Spectral energy distributions of the \gr emission from Tycho's SNR
(total spectrum, which is the sum of all components, (a), and separately
$\pi^0$-decay, inverse Compton (IC) and non-thermal Bremstrahlung (NB)
components, (b)), as functions of \gr energy $\epsilon_{\gamma}$, calculated
for different source distances, together with the upper limits measured by the
{\it HEGRA} \citep[H-CT;][]{aha01} and {\it Whipple} \citep[W;][]{buckley98}
Cherenkov telescopes, the {\it HEGRA AIROBICC} \citep[HA;][]{prahl97} upper
limit, and data presented by {\it SHALON} \citep[S;][]{sini07}. The {\it thin
dot-dot-dashed line} in Fig.\ref{f4}a represents the IC spectrum that is
expected to be produced in the test particle case.}
\label{f4}
\end{figure}

In order to find a constraint on the distance $d$ and the ISM density
$N_\mathrm{H}$ from \gr observations, we compare in Fig.\ref{f4}a the resulting
\gr spectral energy distribution with the {\it HEGRA} and {\it Whipple} upper
limits at TeV energies. The \gr fluxes, calculated in spherical symmetry, have
all been reduced by a factor $f_\mathrm{re} = 0.2$, in order to account for the
selective injection of protons which operates only over $\approx 20 \%$ of the
SNR surface \citep{vbk03}.  It can be seen that all distances $d < 3.3$~kpc are
inconsistent with the {\it HEGRA} data\footnote{Given the considerably higher
sensitivity of {\it HEGRA}, the {\it SHALON} results contradict the H-CT
result. Therefore we shall not discuss the claimed detection by {\it
SHALON}.}. We also note that at all distances considered the expected \gr flux
is dominated by $\pi^0$-decay (hadronic) \grs (see Table~1 and
Fig.\ref{f4}b). However, the contribution of these hadronic \grs progressively
decreases with the increase of the source distance: the hadronic \gr flux
$F_{\gamma}^{\pi}\propto E_\mathrm{c} N_\mathrm{H}/d^2$ is proportional to the
ISM number density $N_\mathrm{H}$ and to the total CR energy content
$E_\mathrm{c}$ \citep[e.g.][]{bv97}. In the Sedov phase $E_\mathrm{c}$ is an
almost distance-independent fraction of $E_\mathrm{sn}$, and
$N_\mathrm{H}\propto d^{-5}$. This gives then $F_{\gamma}^{\pi}\propto d^{-7}$,
since $E_\mathrm{sn}$ is fixed. Such a dependence is roughly consistent with
the results of our calculations given in Fig.\ref{f4}.

Fig.\ref{f4}b shows that for $d\geq 4$~kpc the hadronic \gr flux approaches the
inverse Compton (IC) flux.  The latter is almost insensitive to the source
distance $d$. Therefore its value $\epsilon_{\gamma}F_{\gamma}\approx
10^{-2}$~eV/(cm$^2$s) at $10^8< \epsilon_{\gamma}< 10^{13}$~eV represents the
lowest possible limit. The contribution of non-thermal bremsstrahlung (NB)
emission is negligibly small.

\subsection{Simple IC emission model}

The closeness of the predicted hadronic and IC \gr fluxes might at first sight
also be seen to justify the simple consideration that a situation is possible
or even likely in which no nuclear particles are accelerated at all, but only
electrons. The observed upper limit on the \gr flux might then be explained by
the IC \gr emission in the cosmic microwave background (CMB) radiation field of
the same accelerated electrons which also produce the observed synchrotron
emission.

A simple estimate of the expected IC TeV-emission can be obtained from the
relation $\epsilon_{\gamma} F_{\gamma}^{IC}= \nu
S_{\nu}w_\mathrm{ph}/(B_\mathrm{d}^2/8\pi)$ which is valid for the gamma ray
energy $\epsilon_{\gamma}= 1.5\times 10^{10}(h\nu/\mbox{eV})
(10\mu\mbox{G}/B_\mathrm{d})$~eV assuming that IC scattering takes place in the
Thompson regime.  Here $w_\mathrm{ph}\approx 0.3$~eV/cm$^3$ is the CMB energy
density.  Since according to Fig.\ref{f2} we have $\nu S_{\nu}\approx
100$~eV/(cm$^2$s) for $h\nu =1$~keV ($\nu \approx 2\times 10^{17}$~Hz), the
expected IC \gr energy flux at $\epsilon_{\gamma}=15
(10\mu\mathrm{G}/B_\mathrm{d})$~TeV produced by the same electrons is
$\epsilon_{\gamma} F_{\gamma}^{IC}\approx 10(10 \mu
G/B_\mathrm{d})^2$~eV/(cm$^2$s).  These relations show that the IC \gr flux
exceeds the HEGRA upper limit unless the interior magnetic field is considerably
smaller than 10~$\mu$G.

In order to illustrate the energy spectrum expected in the case of inefficient
proton acceleration we present in Fig.4a a IC \gr spectrum, calculated at
$B_\mathrm{d}=20$~$\mu$G an with electron energy distribution function
$f_\mathrm{e}(\epsilon)=A\epsilon^{-2}\exp(-\epsilon/\epsilon_\mathrm{max})$,
which corresponds to the test particle solution. The parameters $A$ and
$\epsilon_\mathrm{max} = 13$~TeV in this expression were determined by fitting
the observed synchrotron flux $\nu S_{\nu}$ at $\nu=1$~GHz and $\nu=3\times
10^{18}$~Hz (see Fig.\ref{f2} and \ref{f4}).

As expected from the results of the nonlinear theory, this simple test particle
theory which considers an inefficient production of nuclear CRs is quite
inconsistent with the existing upper limit for the \gr production. Since the
synchrotron emission is expected to come from those regions of the shock
surface where the field is strongly amplified, the shock is a quasi-parallel
shock. Therefore the ISM magnetic field which
should have a strength of $5\mu$G or less, should be MHD-compressed by a factor
that is significantly less than the shock compression ratio. As a result an
interior field strength $B_\mathrm{d}$ of about $10\mu$G is adequate for this
type Ia SN. In Fig.\ref{f4} a field strength $B_\mathrm{d}=20\mu$G was
chosen. It still overpredicts the \gr flux by an order of magnitude. Only for
$B_\mathrm{d} \approx 70\mu$G the predicted IC \gr flux would be as low as the
observed HEGRA upper limit\footnote{We remark that a similar result is found
for SN1006, the other prototypical type Ia SN observed, where for $B_\mathrm{d}
\approx 20\mu$G the analogously calculated IC flux equals the H.E.S.S. upper
limit \citep{aha05} for the \gr flux. Any smaller field $B_\mathrm{d} < 20\mu$G
would overpredict the \gr flux.}. This result shows that the assumption of
electron acceleration alone in SNRs has no physical basis and should be
abandoned.

\subsection{Interior magnetic field strength from X-ray filaments}

We have also determined the interior magnetic field value from the observed
linear thickness $L$ of X-ray filaments \citep[see][for
details]{vbk05, vbk05a}. These values are given in Table 1 as $B'_\mathrm{d}$
for different distances. Since the size $L$ increases proportional to $d$, the
magnetic field strength $B'_\mathrm{d}\propto L^{-2/3}$ \citep{bv04b} decreases
with increasing $d$. Therefore there is a growing discrepancy in the ratios between
$B_\mathrm{d}$ and $B'_\mathrm{d}$: at $d=4.5$~kpc $B'_\mathrm{d}\approx
310\mu$G which is already considerably smaller than $B_\mathrm{d}\approx
370\mu$G. $B'_\mathrm{d}/B_\mathrm{d}$ is already considerably smaller than at
$d=3.1$~kpc. Therefore we might be able to
constrain the source distance also from above, $d<4$~kpc.

\section{Conclusions}
The last result led \citet{vbk07} to the conclusion that one could constrain
the source distance also from above, $d < 4$~kpc. However, this is not a strong
argument, as the detailed data presented in Table~1 show. Therefore, regarding
the expected \gr flux, we can only say the following: the increase in the
kinetic energy of explosion to $E_\mathrm{sn} = 1.2 \times 10^{51}$~erg
requires a rather high source distance of $d > 3.3$~kpc to explain the HEGRA
non-detection. On the other hand, the rather low distance estimates from
independent measurements summarized in the beginning of this section
make it for this increased value of $E_\mathrm{sn}$ even more likely that the
actual \gr flux from Tycho is ``only slightly'' below the HEGRA upper
limit. The strong magnetic field amplification implies that the \gr flux is
hadronically dominated, at least above photon energies of 1 GeV.

From the agreement between our theoretical solutions with the measurements of
the discontinuity radii we conclude that there is new evidence for effective
acceleration of nuclear particles in Tycho's SNR. A future \gr detection would
therefore make the case for {\it nuclear} particle acceleration in SNRs
incontrovertible.

\begin{acknowledgements}
HJV is indebted to Wolfgang Hillebrandt for discussions about SN explosion
energies and to Roger Chevalier for a discussion on the modification
of the contact discontinuity by the Rayleigh-Taylor instability. EGB and LTK
acknowledge the partial support by the Presidium of RAS (program No.16) and by
the Russian Foundation for Basic Research (grant 07-02-00221) and the
hospitality of the Max-Planck-Institut f\"ur Kernphysik, where part of this
work was carried out.
\end{acknowledgements}

\end{document}